\documentclass[prb, twocolumn,superscriptaddress,floatfix]{revtex4-1}
\usepackage{amsmath,amssymb,graphicx,bm,epsfig}
\usepackage{epstopdf}
\usepackage{color}
\newcommand{\vk}{{\mathbf{k}}}

\begin{document}

\title{DFT+DMFT study of dopant effect in a heavy fermion compound CeCoIn$_5$}

\author{Hong Chul Choi}
\email{chhchl@gmail.com}
\affiliation{Theoretical Division, Los Alamos National Laboratory, Los Alamos, New Mexico 87545, USA}
\affiliation{Center for Correlated Electron Systems, Institute for Basic Science (IBS), Seoul 08826, Korea}

\author{Eric D.~Bauer}
\affiliation{Materials Physics and Application Division, Los Alamos National Laboratory, Los Alamos, New Mexico 87545, USA}

\author{Filip Ronning}
\affiliation{Institute for Materials Science, Los Alamos National Laboratory, Los Alamos, New Mexico 87545, USA}

\author{Jian-Xin Zhu}
\email{jxzhu@lanl.gov}
\affiliation{Theoretical Division, Los Alamos National Laboratory, Los Alamos, New Mexico 87545, USA}
\affiliation{Center for Integrated Nanotechnologies, Los Alamos National Laboratory, Los Alamos, New Mexico 87545, USA}

\begin{abstract}
We study the dopant-induced inhomogeneity effect on the electronic properties of  heavy fermion CeCoIn$_5$ using a combined approach of density functional theory (DFT) and dynamical mean-field theory (DMFT). 
The inhomogeneity of the hybridization between Ce-$4f$  and conduction electrons is introduced to impose the inequivalent Ce atoms with respect to the dopant.
From the DFT to the DFT+DMFT results, we demonstrate a variation of the hybridization strength depending on the hole or electron doping.
A drastic asymmetric mass renormalization could be reproduced  in the DFT+DMFT calculation.
Finally, the calculated coherence temperature reflects the different development of the heavy quasiparticle states, depending on the dopant.
\end{abstract}
	
%Physics Subject Headings
%Heavy-fermion systems
%Kondo effect
%dynamical mean field theory
%dopant

\date{\today}
\maketitle
\section{Introduction}
 Heavy fermion systems (HFSs)\cite{Fulde:2012,Fulde:2006,Pfleiderer:2009} have shown unconventional superconductivity, Kondo effect, valence fluctuations, magnetism and  exotic coexisting phases.
 The renormalized quasiparticle kinetic energies observed in the heavy fermion originates from the Kondo effect.
 The strength of the Kondo effect is determined by the hybridization function ($\Delta$) between localized and conduction electrons,  and the density of state ($N_F$) at the Fermi level ($E_F$).
Strong hybridization could suppress the renormalization effect over the competition with the Coulomb interaction.
 The wavefunction of the $4f$-electrons is deeply distributed inside the atomic radius close to the atomic wavefunction. 
 However,  the energy gain through the Kondo effect could drive a crossover from a localized moment into a highly renormalized  itinerant quasiparticle state near $E_F$. 
 In addition, the atomic multiplet of Ce-4$f$ valence states provide an additional incoherent feature, being identified as lower and upper Hubbard bands. The three peak spectral function $A(k,\omega)$ represents the characteristic dual nature of  the strongly correlated $f$-electron system.

%{\bf Introduction of CeTIn5}
The Ce-based heavy fermion compounds are one prototypical family of HFSs. CeIn$_3$ shows an antiferromagnetic ground state with $T_N$=10 K, where superconductivity with $T_c$=0.21 K emerges under a critical pressure of 2.6 GPa.~\cite{Mathur:1998,Knebel:2001} 
CeMIn$_5$ (M=Co, Rh, Ir)\cite{Petrovic:2001a,Hegger:2000,Petrovic:2001b} is synthesized by inserting MIn$_2$ layers between CeIn$_3$ layers. 
In these compounds, CeCoIn$_5$  and CeIrIn$_5$ are unconventional superconductors at $T_c$=2.3  K and 0.4 K, respectively, indicative of Ce-4$f$ electrons being delocalized, whereas CeRhIn$_5$ is an antiferromagnet ($T_N$=3.8 K) with localized 4$f$ electrons at ambient pressure. 
 The substitution of  Cd~\cite{Pham:2006,Nicklas:2007,Urbano:2007,Rusz:2008,Booth:2009,Gofryk:2012,Sakai:2017,Chen:2019},
 Ru~\cite{Ru:2015}
 or Hg~\cite{Booth:2009,Gofryk:2012}~(hole doping) and Sn~\cite{Bauer:2005,Bauer:2005b,Bauer:2006,Daniel:2005,Rusz:2008,Booth:2009,Gofryk:2012,Sakai:2017},
 Zn~\cite{Zn:2014,Zn:2020}
 or Pt~\cite{Gofryk:2012}~(electron doping) 
 has been reported to either promote antiferromagnetic (AFM) or metallic phases.
In comparison to In ($4d^{10}5p^{1}$), Cd and Sn have the valence states of $4d^{10}$ and $4d^{10}5p^{2}$, respectively.
The one less (more) electron would be expected to reduce (enhance) the hybridization. 
Experimentally, angle-resolved photoemission spectroscopy has indeed demonstrated  the suppression of the band hybridization in 
Cd-doped CeCoIn$_5$.~\cite{Chen:2019}
The electron or hole doping affects the ground state of CeCoIn$_5$ asymmetrically.
Only one percent Cd substitution causes the AFM ground state with a local moment on the Ce sites ($\sim 0.7\;\mu_{B}$/Ce); while
the Sn substitution leads to a complete disappearance of superconductivity for a critical concentration  $\sim$ $\sim$ 3.6$\%$ Sn.
In addition, the extended X-ray absorption fine structure measurements has suggested the preference of dopant atoms to the Ce-In plane,~\cite{Daniel:2005,Booth:2009} which means an inhomogeneous distribution of dopants.

%The site dependence of the  substitution  was in good agreement with  the density functional theory (DFT) calculation.~\cite{Rusz:2008,Sakai:2015}
Density functional theory (DFT) has proven to be very effective in helping to unravel the effects of doping in the 115s. For example, DFT could help identify various local atomic environments observed by NMR.~\cite{Rusz:2008,Sakai:2015}
DFT calculations also helped to uncover the local increase (decrease) in the hybridization caused by Sn (Cd) dopants.~\cite{Gofryk:2012} 
Furthermore, the open-core (setting  $4f$ states inside the core state) or the conventional DFT (setting $4f$ states as valence states) provides a hint of 
whether $4f$ states might be localized or delocalized.\cite{Rusz:2008} 
Therefore, DFT proved itself as a powerful tool to incorporate material-specific information, that is, the structure and atomic configuration of the real compound. On the other hand, it is known to underestimate the effect of electronic correlations  in heavy fermion systems.
This inadequacy can be remedied by dynamical mean-field theory (DMFT), which has successfully captured the correlation-induced crossover between localization and delocalization  in strongly correlated systems.
Via a strong band renormalization, 
 the correlation effect on the electronic structure and optical properties
 were well reproduced by DFT+DMFT calculations.\cite{Shim:2007,Choi:2012,Nomoto:2014}
Using the virtual crysal approximation,
a DFT+DMFT study reproduced the general trends of the doping effect.\cite{Seo:2015}
However,  the impact of the dopant-induced inhomogeneity  in real compounds has not been addressed.

Here we revisit the effect of  doping on CeCoIn$_5$  to take into account the inhomogeneity with respect to the distance from a dopant to a Ce atom.
We break the equivalence of Ce atoms by placing the dopant close to a specific Ce atom.
The electronic structure is investigated by DFT and DFT+DMFT calculations.
The DFT results give a hint about the different behavior of hole versus electron substitutions, while its limitation is also shown.
The small change in the hybridization in the DFT quasiparticle could turn into  a totally different phase through the DMFT impurity solver.
The inhomogeneity  of the self-energy and the hybridization functions is shown in the DFT+DMFT results.
The different evolution of of the heavy quasiparticle band induced by the dopant will be discussed with the calculated coherence temperature.

%\begin{itemize}
%	\item Why Cd and Sn show different effect ( hole vs electron)
%	\item Why Cd doping effect emerges substantially at even small doping concentration
%\end{itemize}

The paper is organized as follows: Section \ref{Method} explains  the computational method. In  Sec. \ref{Results}  we address calculated crystal structures, the DFT and DFT+DMFT results. Also, the calculated coherence temperature will be provided here. Section \ref{Summary} presents a summary and concluding remarks.  

\section{Computational details} \label{Method}
The charge self-consistent version of DFT+DMFT,~\cite{Kotliar:2006} as implemented in Ref.~\onlinecite{Haule:2010}, is
based on the full-potential linearized augmented plane-wave (FP-LAPW) band method.~\cite{wien2k}
The correlated $4f$-electrons are treated dynamically by the DMFT local self-energy ($\Sigma(\omega)$), 
while all other delocalized $spd$ electrons are treated on the DFT level.
The charge and spin fluctuations considered in DMFT enable the describtion of the Kondo effect correctly.
$\Sigma(\omega)$ is calculated from
the corresponding impurity problem, in which  
full atomic interaction matrix is taken into account ($F^0$= 5.0 eV,
$F^2$= 8.10693641618 eV
$F^4$= 5.41543352601 eV
and, $F^6$= 4.0048265896).~\cite{Cowan}
A  temperature of 5 meV is used in the calculations.
To solve the impurity problem, 
we used the one-crossing approximation of the local self-energies.~\cite{Kotliar:2006}
RK$_{max}$ is 7.0 and 3000 k-point mesh are used through the calculation. 
The spin-orbit coupling is always included in the calculations.
The total charge density and the self-energy through the DFT+DMFT iterations are converged within 10$^{-5}$ and 10$^{-4}$, respectively.

The single-site DMFT solver for the $4f$ state in each unique Ce cite are treated separately.
The $4f$ state for the unique Ce sites will be denoted as Ce1 or Ce2.
The inter-site correlation beyond the single-site DMFT solver is included through the hybridization function ($\Delta$).
The self-consistent equation between the local Green's function  ($\mathrm{G_{loc}}=\frac{1}{\omega-\epsilon_{imp}-\Sigma-\Delta}$, $\epsilon_{imp}$: impurity energy level, $\Sigma$:  the self-energy,$\Delta$: the hybridization function between impurity and bath) and the lattice green function computed by the DFT eigenvalue ($H_\vk^{DFT}$) and $E_F$,
 is defined through the projection operator ($P$) from the lattice to the impurity and the embedding operator ($E$) vice versa.
Using $\hat{P}_{1,\vk}$, $\hat{P}_{2,\vk}$, $\hat{E}_{1,\vk}$, and $\hat{E}_{2,\vk}$   (1:Ce1, 2:Ce2),
the self-consistent equations with the diagonal basis are defined  as follows:
\begin{multline}
\frac{1}{\omega-\epsilon_{1,imp}-\Sigma_{1}-\Delta_{1}}= \\ \sum_\vk \hat{P}_{1,\vk}[(\omega+E_F-H_\vk^{DFT}-\hat{E}_{1,\vk}\Sigma_{1})^{-1}],
\end{multline}
and
\begin{multline}
\frac{1}{\omega-\epsilon_{2,imp}-\Sigma_{2}-\Delta_{2}}= \\ \sum_\vk \hat{P}_{2,\vk}[(\omega+E_F-H_\vk^{DFT}-\hat{E}_{2,\vk}\Sigma_{2})^{-1}].
\end{multline}
Here ,$\epsilon_{i,imp}$, $\Sigma_{i}$, and $\Delta_{i}$ mean the impurity energy level, self-energy, and hybridization function, respectively. $i$ represents the different impurity (Ce).
The long-range correlation effect is taken into account through $\Delta_{i}$.

\begin{figure}[t]
\includegraphics[width=1.0\linewidth,angle=0]{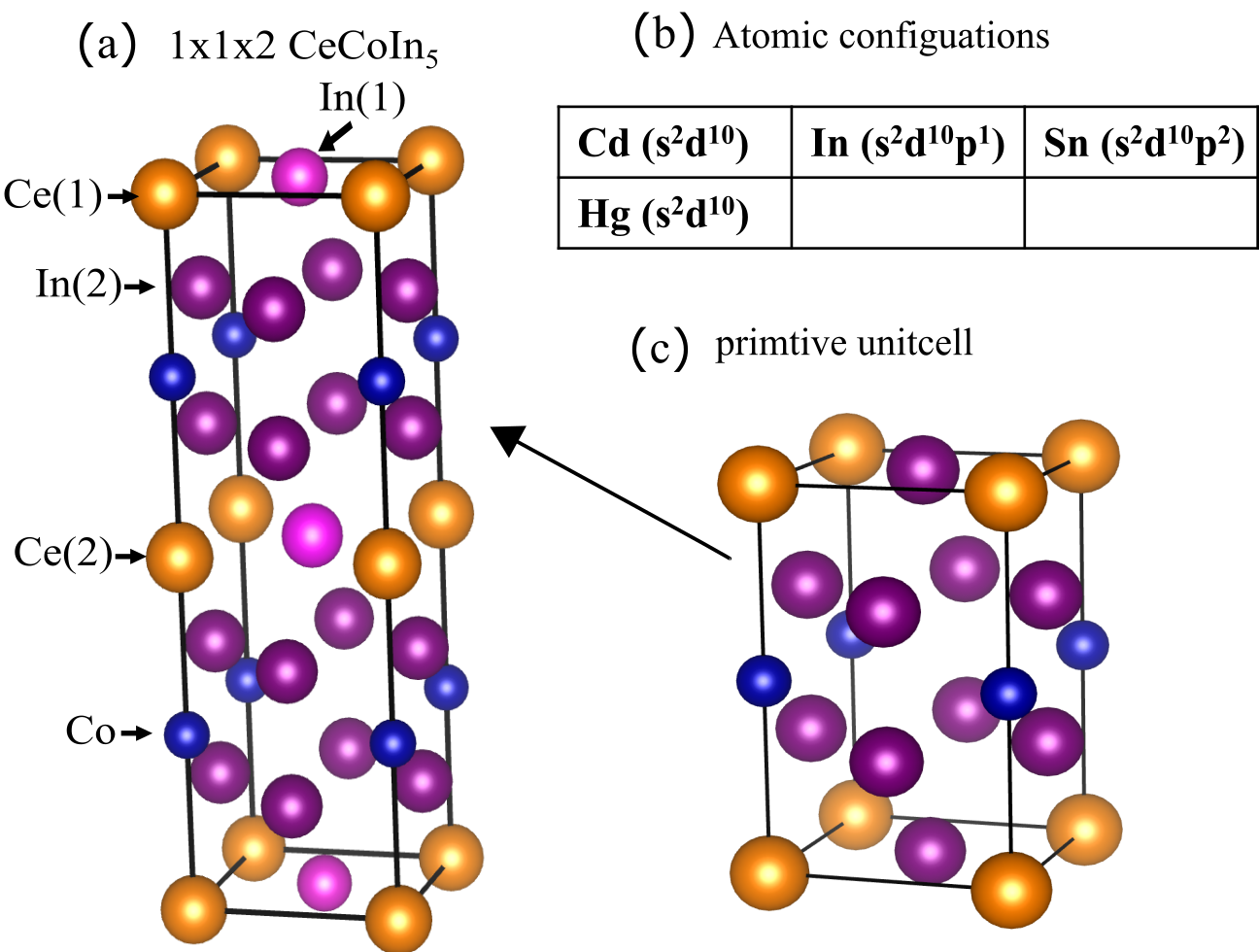}
\caption{(color online) (a)The schematic crystal structures\cite{Vesta} of the 1x1x2 supercell of CeCoIn$_5$  The different colors represent the different atoms. Indium (In) atoms have the magenta (In1) and purple (In0) colors.  For the doped case,  the dopant atom replaces In1 to give the concentration of  10$\%$-doped structure. Additionally, the two-types of Ce atoms are distinguished to Ce1 (close to the dopant) and Ce2 (away from the dopant). (b) The atomic configurations of Cd, In, Sn, and Hg are compared.  (c) The primitive unit cell of CeCoIn$_5$ is provided. 
}
\label{fig1a}
\end{figure}

\begin{figure}[t]
\includegraphics[width=1.0\linewidth,angle=0]{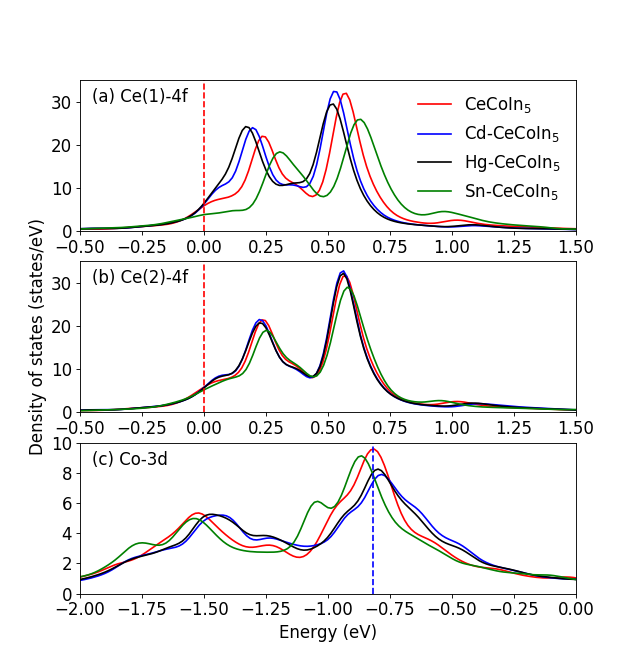}
\caption{(color online)  The density of states for (a) Ce1-4$f$, (b) Ce2 -4$f$, and (c) Co-3$d$ states calculated in four different structures with DFT. The labels represent  CeCoIn$_5$ (pristine), CeCo(In$_{0.9}$Cd$_{0.1}$)$_5$  (Cd-doped CeCoIn$_5$), CeCo(In$_{0.9}$Hg$_{0.1}$)$_5$ (Hg-doped CeCoIn$_5$),  and CeCo(In$_{0.9}$Sn$_{0.1}$)$_5$ (Sn-doped CeCoIn$_5$). The red vertical line represents the Fermi level. The blue vertical line is provided as a guideline to compare the shift of the bands with respect to the dopant.
}
\label{fig1b}
\end{figure}

\section{Numerical Results}\label{Results}
Figure~\ref{fig1a}(a) presents the supercell structure of CeCoIn$_5$  made from the original primitive unit cell as shown Fig.~\ref{fig1a}(c).
We adopt the minimum size of the 1x1x2 supercell of CeCoIn$_5$, which has two Ce,  two Co  and ten In atoms for our calculations. 
 Ce(1) at (0,0,0) and Ce(2) at (0,0,0.5) are equivalent in the pristine structure.
 In the primitive cell, In atoms are distinguished as the ones (In(1)) in the Ce plane and the other four In atoms (In(2)) 
 in between the Co plane and the Ce-In plane.
For simplicity, we choose the dopant atom (Cd,Sn or Hg) to occur at the In(1) position of the Ce(1) plane. This is consistent with extended X-ray absorption fine structure measurement, which indicate that these dopant atoms preferentially substitute on the In(1) site.
 These substitutions force Ce(1) and Ce(2) to become  inequivalent.
In this work, we examine the electronic structure of CeCoIn$_5$ (pristine),  CeCo(In$_{0.9}$Cd$_{0.1}$)$_5$  (Cd-doped CeCoIn$_5$), CeCo(In$_{0.9}$Hg$_{0.1}$)$_5$ (Hg-doped CeCoIn$_5$),  and CeCo(In$_{0.9}$Sn$_{0.1}$)$_5$ (Sn-doped CeCoIn$_5$).

First we report the conventional DFT calculations of the doped CeCoIn$_5$.
Figure~\ref{fig1b} shows the density of states (DOS) of (a) Ce(1)-$4f$ states, (b) Ce(2)-$4f$ states, and (c) Co-$3d$ states. We note that the DOS for the two Co atoms are identical due to the inversion symmetry.
Depending on the dopant, the Ce-$4f$ bands are shifted with respect to the pristine case (red line).
The two peaks in Fig. \ref{fig1b}(a-b) correspond $j$=5/2 (left peak) and $j$= 7/2 (right peak) states split by spin-orbit coupling ($\sim$ 0.3 eV).
This value is inherited into the impurity solver in the DFT+DMFT calculation. 
Hereafter we focus on the  $j$=5/2 states, which mainly contribute the states around $E_F$.

As  Cd and Hg (Sn) have one less (more) electron than In (See Fig.~\ref{fig1a}(b)), 
the shift of the chemical potential with respect to  Ce-$4f$ states would be lowered (increased) in the Cd- and Hg-doped (Sn-doped) cases.  
The trend in Fig.~\ref{fig1b}(a-b) for Ce(1)-$4f$ and Ce(2)-$4f$ in the pristine and doped cases does not follow this simple intuition.
In comparison to the $j$=5/2 peak of the pristine Ce-$4f$ states,
the peak for Cd- and Hg-doped (Sn-doped) cases is shifted toward (away from) the Fermi energy level.
Also,  the DOS at $E_F$ are increased in Cd- and Hg-doped CeCoIn$_5$ systems, whereas that of the Sn-doped CeCoIn$_5$  is decreased.
Those results indicate the effect of doping on the $4f$ states could not be  described well by the DFT method.
On the other hand, the inhomogeneous feature between Ce(1) and Ce(2) sites could be identified.

We also explore the doping effects on the 3$d$-bands of transition-metal Co.
As shown in  Fig. \ref{fig1b}(c),
the band center of Co-3$d$ states is shifted toward $E_F$ in the Cd- and Hg-doped CeCoIn$_5$ cases, while  that of the Sn-doped CeCoIn$_5$  is moved away from $E_F$. Such behavior is in agreement with a rigid band picture.

\begin{figure}[t]
	\includegraphics[width=1.0\linewidth,angle=0]{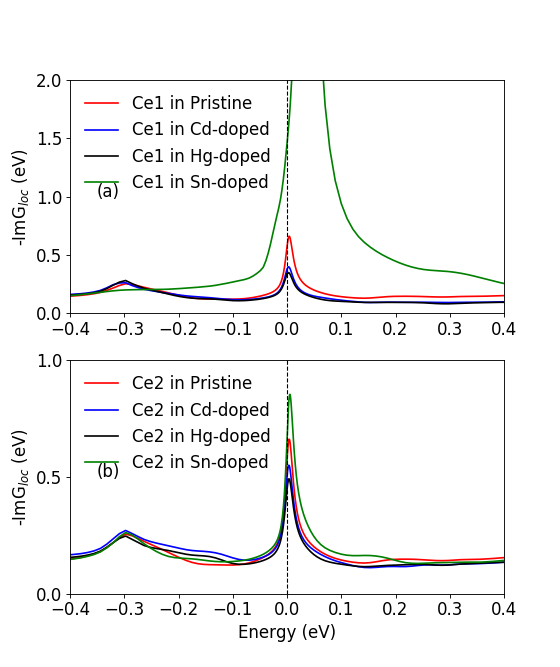}
	\caption{(color online)
				The imaginary part of the local Green's functions for (a) Ce1 and (b) Ce2 in different structures with DFT+DMFT.The labels represent  CeCoIn$_5$ (Pristine), CeCo(In$_{0.9}$Cd$_{0.1}$)$_5$, CeCo(In$_{0.9}$Hg$_{0.1}$)$_5$,  and CeCo(In$_{0.9}$Sn$_{0.1}$)$_5$.
	}
	\label{fig_ImG}
\end{figure}

In the remainder, we will discuss  how correlations influence the electron- and hole-doped cases.
Through the DFT+DMFT calculations, Ce(1)-$4f$ and Ce(2)-$4f$ states are treated as  correlated impurity states,  denoted by Ce1 and Ce2.
The DOS ($-\mathrm{Im}\mathrm{G_{loc}}$) of Ce1 and Ce2 are provided in Fig.~\ref{fig_ImG}.
In CeCoIn$_5$, the  peak of $j$=5/2 states grows as temperature lowers.
Fig.~\ref{fig_ImG} is a snapshot  in the middle of the development of the quasiparticle peak.
In comparison to Fig.~\ref{fig1b} (a,b), 
the width of the peaks are highly reduced and pushed towards  $E_F$.
The DOS at $E_F$ of  the Ce1 in the Sn-doped case is much larger than others.
Also no spin-orbit side band at $-0.3$ eV could be observed. 
The spin-orbit side band is attributed to the transition between $j$=5/2 and $j$=7/2 states.
The Sn dopant due to the increased hybridizations pushes the Ce1 away from the Kondo limit.
The enhanced DOS of the Ce1 is associated with its closer proximity to the Fermi liquid in CeCo(In$_{0.9}$Sn$_{0.1}$)$_5$. 
In case of the hole-doped calculations (Cd or Hg), Ce1 shows a reduction of the DOS relative to the pristine case.
This opposition behavior is consistent with stabilizing a magnetic state upon hole doping.

\begin{figure}[t]
	\includegraphics[width=1.0\linewidth,angle=0]{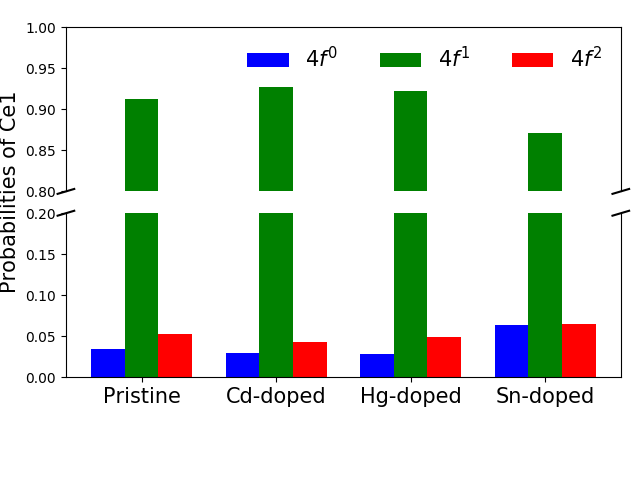}
	\caption{(color online)
		The calculated probabilities of valence states for $n_f$=0 (blue), 1 (green), and 2 (red) of Ce1 in the pristine and doped cases.
	}
	\label{fig_prob}
\end{figure}

We examine the $4f$ electron occupancy ($n_f$) for the pristine and doped cases. 
As summarized in Table~\ref{nf-table},
the calculated $n_f$ in Sn-doped case is larger than the others.
This trend looks contradicting to the intuitive view of valence change from $4f^1$ ($\mathrm{Ce}^{3+}$) to $f^{0}$ ($\mathrm{Ce}^{4+}$) under strong hybridization. Therefore, within a simple hybridization picture, one would expect that the hole (electron) doping would drive the valence of Ce states toward $f^{1}$ ($f^{0}$).
Here we bring up a new insight:  As shown in Fig.~\ref{fig_prob}, although the total number of Ce-4$f$ electron occupation is larger in the Sn-doped case,  under the electron doping,  the probability of $f^{1}$ configuration decreases while simultaneously the probabilities of $f^{0}$ and $f^{2}$ configurations increase.
We thus conclude that  the calculated $n_f$ measures both the Ce-4$f$ electron count for both low energy and high energy configuration states.
The overall differences of $n_f$ are less than 0.1 electron per Ce atom.
However, the correlation effect enables the fine tuned electron number to bring up the drastic change of its electronic structure.

 \begin{table}
	\caption{Ce-$4f$ electron occupancy, $n_f$, and probabilities of $4f^0$, $4f^1$, and $f^2$ for Ce1 in the DFT+DMFT calculations.}
    \begin{tabular}{ | l | l |  l | l| l| l|}
    \hline
     n$_f$ (electron/Ce atom)  & Ce1 & Ce2 &  P($4f^0$)&  P($4f^1$)&  P($4f^2$)\\ \hline
    CeCoIn$_5$  & 1.0432 & 1.0432  & 0.0341 & 0.9126 &0.0532     \\ \hline
    CeCo(In$_{0.9}$Cd$_{0.1}$)$_5$ & 1.0299 & 1.0329 &0.0297 &0.9267&0.0434       \\ \hline
    CeCo(In$_{0.9}$Hg$_{0.1}$)$_5$ & 1.0214 & 1.0362 &0.028 &0.9224 & 0.0495        \\ \hline
    CeCo(In$_{0.9}$Sn$_{0.1}$)$_5$ & 1.0822 & 1.0452 &0.064&0.8711&0.0646  \\
    \hline
    \end{tabular}
   \label{nf-table}
\end{table}

\begin{figure}[t]
	\includegraphics[width=1.0\linewidth,angle=0]{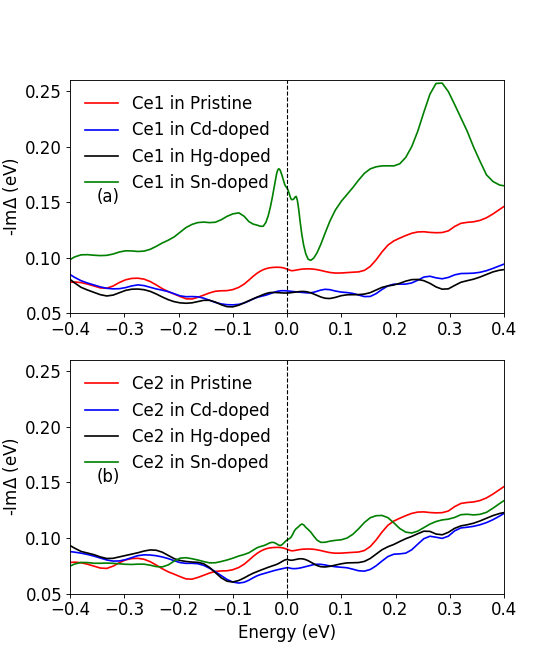}
	\caption{(color online)
		The imaginary part of the hybridization functions for (a) Ce1 and (b) Ce2 in different structures with DFT+DMFT. The labels represent  CeCoIn$_5$ (Pristine), CeCo(In$_{0.9}$Cd$_{0.1}$)$_5$, CeCo(In$_{0.9}$Hg$_{0.1}$)$_5$,  and CeCo(In$_{0.9}$Sn$_{0.1}$)$_5$.
	}
	\label{fig_dlt}
\end{figure}

 \begin{table}
	\caption{Summary of the hybridization strength $-\mathrm{Im}\Delta(0)$.}
    \begin{tabular}{ | l | l|l| l |  l |}
    \hline
    $-\mathrm{Im}\Delta(0)$ (eV)   & Ce1 & Ce2 \\ \hline
    CeCoIn$_5$                    &  0.08999 & 0.08999      \\ \hline
    CeCo(In$_{0.9}$Cd$_{0.1}$)$_5$ & 0.07032 & 0.07345       \\ \hline
    CeCo(In$_{0.9}$Hg$_{0.1}$)$_5$ & 0.06837 & 0.08064       \\ \hline
    CeCo(In$_{0.9}$Sn$_{0.1}$)$_5$ & 0.16292 & 0.0982  \\
    \hline
    \end{tabular}
   \label{Delta-table}
\end{table}

We investigate the calculated hybridization function ($-\mathrm{Im}\Delta$ ) for Ce1 and  Ce2 as shown Fig.~\ref{fig_dlt}(a,b).
Compared to Fig.~\ref{fig1b}(a-b), the DFT+DMFT calculation  shows the hybridization strength increases with the number of valence electrons. 
In energy between $-0.1$ eV and 0.1 eV,  the  dopant-dependent hybridization functions are well illustrated with respect to the red lines (pristine).
%The values at $E_F$ are summarized in Table~\ref{Delta}.
The imaginary part of the hybridization ($-\mathrm{Im}\Delta (\omega=0)$) at E$_F$ of Ce1 in  Sn-doped CeCoIn$_5$ is enhanced significantly over the other cases (see Table~\ref{Delta-table}). 
This enhancement of hybridization is reduced to be comparable to the pristine case for Ce2. 
This could be understood in terms of  the relative distances between the Ce atoms and the Sn atom. 
The substitution of a hole (Cd, Hg) decreases $-\mathrm{Im}\Delta (\omega=0)$ around $E_F$.
Both Cd and Hg dopants give rise to similar decreases of the  $-\mathrm{Im}\Delta (\omega=0)$ on both Ce1 and Ce2 with respect to the pristine cases.
The behavior shown in $-\mathrm{Im}{\Delta}$ and $-\mathrm{Im}\mathrm{G_{loc}}$ could reveal how the hybridization is changed in each case. 
 The reduction of the hybridization favors a local-moment ground state over the Fermi liquid phase.
 We conclude that  Sn-doped ( Cd- or Hg-doped) CeCoIn$_5$ would be in proximity to a Fermi liquid state (a magnetic ground state).

\begin{figure}[t]
	\includegraphics[width=1.0\linewidth,angle=0]{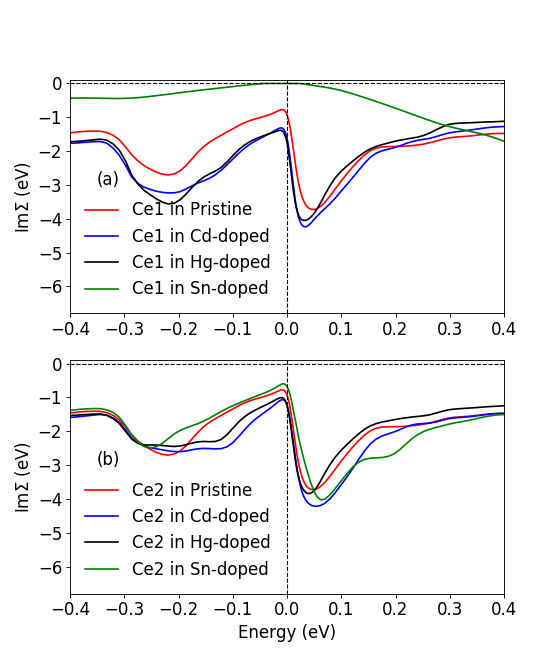}
	\caption{(color online)
			The imaginary part of the self-energy for (a) Ce1 and (b) Ce2 in different structures with DFT+DMFT.  The labels represent  CeCoIn$_5$ (Pristine), CeCo(In$_{0.9}$Cd$_{0.1}$)$_5$, CeCo(In$_{0.9}$Hg$_{0.1}$)$_5$,  and CeCo(In$_{0.9}$Sn$_{0.1}$)$_5$. 
	}
	\label{fig_ImS}
\end{figure}

To elaborate more insight into the renormalization of the quasiparticle state, we will examine the imaginary ($\mathrm{Im}$) and real ($\mathrm{Re}$) parts of the calculated self-energy ($\Sigma$).
Figure~\ref{fig_ImS} shows  the imaginary part of the self-energy ($\mathrm{Im}\Sigma$),whose inverse  at $E_F$ represents the inverse of the life-time of quasiparticle state. 
$\mathrm{Im}\Sigma (\omega=0)$ of  Ce1 in the Sn-doped case is almost zero at $T$=0.005 eV ($\sim$ 65 K).
The parabolic behavior of $\mathrm{Im} \Sigma$ around $\omega =0$ of Ce1 in Sn-doped CeCoIn$_5$ is a typical Fermi liquid characteristic.
The Fermi liquid phase represents the presence of the coherent quasiparticle states.
The finite values of of $\mathrm{Im} \Sigma$ ($\omega=0$) indicate the deviation from the Fermi liquid phase in the other cases.
This quasiparticle state of the Ce1 in CeCo(In$_{0.9}$Sn$_{0.1}$)$_5$ would be consistent with the largely enhanced hybridization function.

\begin{figure}[t]
	\includegraphics[width=1.0\linewidth,angle=0]{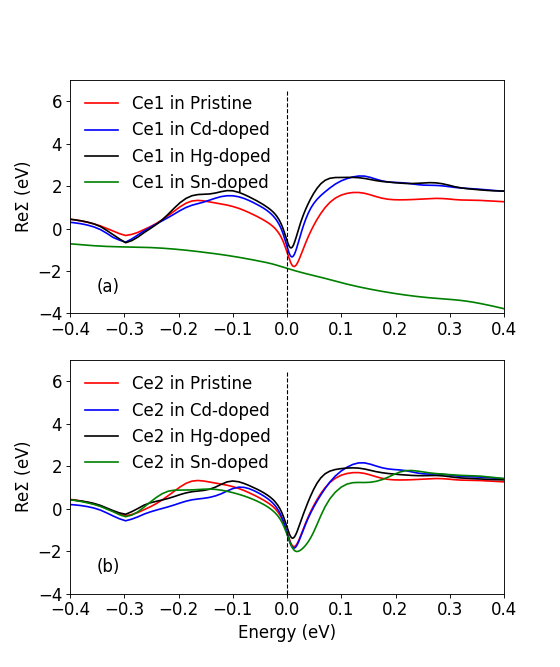}
	\caption{(color online)
			The real part of the self-energy for (a) Ce1 and (b) Ce2 in different structures with DFT+DMFT.	The labels represent  CeCoIn$_5$ (Pristine), CeCo(In$_{0.9}$Cd$_{0.1}$)$_5$, CeCo(In$_{0.9}$Hg$_{0.1}$)$_5$,  and CeCo(In$_{0.9}$Sn$_{0.1}$)$_5$.
	}
	\label{fig_ReS}
\end{figure}

 \begin{table}
	\caption{Summary of the quasiparticle weight ($Z$) at the Fermi energy $E_F$.}
    \begin{tabular}{ | l |l|l| l |  l |}
    \hline
    $Z$ ($1/(1-\frac{\partial \mathrm{Re}\Sigma}{\partial \omega})$ at E$_F$)   & Ce1-$4f$ & Ce2-$4f$ \\ \hline
    CeCoIn$_5$                     & 0.0124  & 0.0124      \\ \hline
    CeCo(In$_{0.9}$Cd$_{0.1}$)$_5$ & 0.0110 & 0.0110       \\ \hline
    CeCo(In$_{0.9}$Hg$_{0.1}$)$_5$ & 0.0121 & 0.0121       \\ \hline
    CeCo(In$_{0.9}$Sn$_{0.1}$)$_5$ & 0.1100 & 0.0139  \\
    \hline
    \end{tabular}
   \label{Z-table}
\end{table}

The renormalization of the band is strongly associated with the mass enhancement under correlation.
The slope of $\mathrm{Re}\Sigma$ at $E_F$ is used to represent how large correlation effect reduces the bandwidth of a quasiparticle state.
Figure~\ref{fig_ReS} provides the variation of $\mathrm{Re}\Sigma$ from $-0.4$ eV to 0.4 eV.
The slopes of $\mathrm{Re}\Sigma$ at $E_F$ seem to be similar to the red lines of the pristine case except for that of Ce1 in Sn-doped case. 
The quasiparticle weight $Z$ is computed using $1/(1-\frac{\partial \mathrm{Re}\Sigma}{\partial \omega})$ at $E_F$, and is given in Table~\ref{Z-table}.
The $\mathrm{Re}\Sigma (\omega=0)$ of the Ce1 shows a very different behavior in comparison to the other cases.
More strikingly, the quasiparticle weight $Z$ (0.11) of the Ce1 in the Sn-doped CeCoIn$_5$ is nearly one order of magnitude larger than that for both the pristine and hole-doped cases.
The larger quasiparticle weight represents the larger overlap of the correlate and non-interacting wavefunction at $E_F$.
The weight for the Ce2 is around 0.0139, which is  slightly larger than the other cases (see Fig.~\ref{fig_ReS}(b)).
The inhomogeneity due to the Sn dopant is consistently observed in the DFT+DMFT calculations. 
On the other hand, the quasiparticle weight  for the hole-doped CeCoIn$_5$ is 0.011 (Cd-doped), and 0.012 (Hg-doped), respectively.
This is also consistent with the observation that  the bandwidths of the pristine and the hole-doped CeCoIn$_5$ compounds are similarly renormalized (see Fig.~\ref{fig_ImG}(a)).
The inhomogeneity due to the hole dopant is relatively  suppressed in the DFT+DMFT calculations.

 \begin{table}
	\caption{Calculation of  coherence temperature, $T_{coh}= -\pi Z \mathrm{Im} \Delta(0)/4$.}
    \begin{tabular}{ | l | l|l| l |  l |}
    \hline
    $T_{coh}$ (eV)   & Ce1-$4f$ & Ce2-$4f$ \\ \hline
    CeCoIn$_5$                     & 0.00087 & 0.00087      \\ \hline
    CeCo(In$_{0.9}$Cd$_{0.1}$)$_5$ & 0.000607 & 0.000634       \\ \hline
    CeCo(In$_{0.9}$Hg$_{0.1}$)$_5$ & 0.000649 & 0.000766       \\ \hline
    CeCo(In$_{0.9}$Sn$_{0.1}$)$_5$ & 0.0140 & 0.001071  \\
    \hline
    \end{tabular}
   \label{Tcoh-table}
\end{table}

Using the calculated $Z$ and $\mathrm{Im}\Delta$, the coherence temperatures ($T_{coh}$) could be predicted using the empirical formula ($T_{coh}= -\pi Z  \mathrm{Im} \Delta(0)/4$).~\cite{Zhu:2015:Tc} 
The calculated $T_{coh}$ is summarized in Table~\ref{Tcoh-table}. 
The largest $T_{coh}$ is 0.0140 eV for Ce1 in  CeCo(In$_{0.9}$Sn$_{0.1}$)$_5$, which is one order of magnitude larger than the used temperature ($T$=0.005 eV). 
However, The $T_{coh}$ for Ce2 in CeCo(In$_{0.9}$Sn$_{0.1}$)$_5$ is reduced to 0.001071 eV, which is still
higher than all other coherence temperatures in the hole-doped compounds.
These results suggest  the substantial enhancement effect of the Sn dopant on the Kondo coherence, which pushes the system to be more close to the Fermi liquid phase, and a robust nonequivalence of the Kondo coherence between Ce1 and Ce2  through the DFT+DMFT iterations.
The hole-doped case show the decrease in both the $-\mathrm{Im}\Delta$ and  $Z$ at $E_F$.
Therefore, $T_{coh}$ in the hole-doped cases is lower than those of the pristine case.
Unlike the electron-doped case, Ce1 and Ce2 simultaneously become more incoherent at $T$=0.005 eV.

\section{Concluding Remarks and Summary}\label{Summary}
Since dopants prefer to be substituted in the Ce plane,~\cite{Booth:2009} our study here has considered the spatial inhomogeneity in the doped CeCoIn$_5$ system.
The much enhanced hybridization $\Delta$ and the non Ce-4$f$ density of states $N_F$ of Ce1 near Sn can drive the system deep into  the Fermi liquid state. 
We have also found through the two single-site DMFT impurity solvers that  the 4$f$-electron behavior is dramatically different on Ce1 and Ce2.
Furthermore, the current study has shown a short-ranged electron doping effect.
On the contrary, in the case of hole-doped CeCoIn$_5$, Ce-$4f$ state would be more incoherent than in the pristine.
We further note that our study has been limited to the paramagnetic solutions and as such the presence of a G-type antiferromagnetic configuration ($q=(\pi,\pi,\pi)$), as reported in the Cd-doped CeCoIn$_5$,~\cite{Nicklas:2007}  is beyond the scope of the present work.
However, the dopant induced incoherence obtained here, indicating that the state moves away from the Fermi liquid phase, is consistent with the experiment. 
It is worthwhile to note that while the Sn doping gives a stronger inhomogeneity effect on the local electronic structure, the NMR experiment has observed~\cite{Sakai:2015} that the Cd doping causes a stronger inhomogeneity effect on the magnetic fluctuations.

In summary, we have investigated the inhomogeneous hole and electron doping effect on CeCoIn$_5$ within the DFT+DMFT framework.
%The change in the hybridization function at $E_F$ by various hole and electron rich dopants has been obtained in  both DFT and DFT+DMFT calculations.
We have revealed a clear variation of the hybridization function as well as the highly renormalized bands of Ce-$4f$ states.
 In addition, the calculated $T_{coh}$ has provided a good insight into whether Ce-$4f$ states in a doped CeCoIn$_5$ system would be coherent or incoherent.
 Specifically, we have discovered a highly inhomogeneous influence of Sn substitution on Ce-4$f$ electron states. This prediction should be accessible to scanning tunneling microscopy experiments on a (100)- or (010)-oriented surface of a Sn-doped CeCoIn$_5$.

\begin{acknowledgments}
Hongchul Choi thanks  Soobeom Seo for illuminating discussions. 
This work was carried out under the auspices of the U.S. Department of Energy (DOE) National Nuclear Security Administration (NNSA) under Contract No. 89233218CNA000001.
It was supported by LANL LDRD Program (H.C.C.), the DOE BES "Quantum Fluctuations in Narrow Band Systems" project (E.D.B, F.R), and NNSA Advanced Simulation and Computing Program (J.-X.Z.). Additional support has been provided in part by the Center for Integrated Nanotechnologies, a DOE BES user facility, in partnership with the LANL Institutional Computing Program for computational resources.
H.C.C.  was also supported by Institute for Basic Science in
Korea (IBS-R009-D1).
\end{acknowledgments}

\bibliographystyle{apsrev}
\bibliography{mybiblio}

\end{document}